\documentclass[aps,pra,groupedaddress,twocolumn]{revtex4-2}

\usepackage{graphicx}% Include figure files
\usepackage{dcolumn}% Align table columns on decimal point
\usepackage{amsmath}
\usepackage{amssymb}
\usepackage{bm}% bold math
\usepackage{xcolor}
\begin{document}

%\preprint{APS/123-QED}

%\title{Line shape of coherent-population-trapping resonances in buffer-gas vapor cells\\under counterpropagating circularly polarized light beams:\\Sign reversal and light shift suppression effects}

\title{Total-field atomic magnetometry using an elliptically-polarized\\frequency-modulated light beam for geomagnetic-field applications}

%\title{Coherent population trapping in buffer-gas alkali-metal vapor cells\\under counterpropagating light beams: Line shape and light shift suppression effect}

\author{D.V.~Brazhnikov}
\email[Corresponding author: ]{x-kvant@mail.ru}
\author{A.O.~Makarov}
\author{K.S.~Kozlova}
\author{M.A.~Rusina}
\author{A.M.~Mikhailov}
\author{V.A.~Vasiliev}
%\author{V.I. Vishnyakov}
\author{E.V.~Tanakov}
\author{A.N.~Goncharov}

\affiliation{Institute of Laser Physics SB RAS, 15B Lavrentyev Avenue, Novosibirsk 630090, Russia}

\author{M.A.~Bobrov}
\author{Ya.N.~Kovach}
\author{A.A.~Blokhin}
\author{N.A.~Maleev}
\author{S.A.~Blokhin}

\affiliation{Ioffe Institute RAS, 26 Politekhnicheskaya Street, Saint Petersburg 194021, Russia}

\date{\today}

\begin{abstract}

%We examine an all-optical scheme for high-sensitivity measurements of the Earth's scale magnetic field ($B$). The scheme is based on the Bell--Bloom technique with frequency-modulated light. A single elliptically polarized light wave is used both for pumping the alkali-metal atoms and probing Larmor precession of their spins under the external magnetic field. In contrast to other similar schemes employing elliptically polarized light, in our case the magnetic resonance manifests not in the rotation of the light polarization but rather in a variation of the ellipticity parameter. In the experiments, a vertical-cavity surface-emitting laser is used to irradiate a $0.125$~cm$^3$ cesium vapor cell filled with a buffer gas. The estimated sensitivity of the magnetic field measurements is about $220$~fT/$\surd$Hz at $B$$\,\approx\,$$50$~$\mu$T, while the bandwidth is approximately $1.5$~kHz. The proposed robust single-beam scheme is suitable for miniaturization and is therefore of particular interest for the development of a compact, high-sensitivity, low-power-consumption atomic magnetometer for a wide range of geophysical applications.

We investigate an all-optical scheme for high-sensitivity measurements of the Earth's scale magnetic field ($B$). The scheme is based on the Bell--Bloom technique with frequency-modulated light. A single elliptically polarized light beam is used both for pumping the alkali-metal atoms and probing Larmor precession of their spins under the external magnetic field. In contrast to many other high-sensitivity magnetometry schemes based on nonlinear polarization rotation of the light, the proposed approach enables the observation of magnetic resonance via changes in the light-beam ellipticity parameter. The approach has been validated in experiments with a vertical-cavity surface-emitting laser, irradiating a $0.125$~cm$^3$ cesium vapor cell filled with a buffer gas. In the current experimental conditions, the achieved sensitivity is estimated at $220$~fT/$\surd$Hz under $B$$\,\approx\,$$50$~$\mu$T, while the shot-noise floor corresponds to $\approx\,$$20$~fT/$\surd$Hz. The bandwidth is estimated at $\approx\,$$1$~kHz. The proposed simple and robust single-beam scheme is well suited for miniaturization and is therefore particularly promising for the development of compact, highly sensitive, low-power-consumption atomic magnetometers for a wide range of applications in the geomagnetic field.

%In contrast to other similar schemes employing elliptically polarized light, in our case the magnetic resonance manifests not in the rotation of the light polarization but rather in a variation of the ellipticity parameter. In the experiments, a vertical-cavity surface-emitting laser is used to irradiate a $0.125$~cm$^3$ cesium vapor cell filled with a buffer gas. The estimated sensitivity of the magnetic field measurements is about $220$~fT/$\surd$Hz at $B$$\,\approx\,$$50$~$\mu$T, while the bandwidth is approximately $1.5$~kHz.

\end{abstract}

\maketitle

\section{\label{sec:1}Introduction}

Optically pumped atomic magnetometers (OPMs) are among the most sensitive quantum sensors of magnetic fields \cite{Book2013,Fabricant2023,Kitching2025}. OPMs have found a broad range of applications in both fundamental and applied physics. For instance, these quantum devises are employed in the search for ``new physics'', such as axion-like dark matter \cite{Afach2021}. OPMs are widely implemented in noninvasive medical diagnostics \cite{Nordenstrom2024,Fedosov2025,Schofield2025,Petrenko2026}, mineral exploration \cite{Dang2010,Perikleous2025,Zhou2026} and archaeological prospection using unmanned aerial vehicles (UAVs) \cite{Cao2022,Accomando2024,Yu2025}.

%An overview of the use of drone-borne quantum magnetometers for geophysical applications is provided in Ref. [1].

%For a long time, these tasks relied either on classical sensors, such as fluxgate magnetometers \cite{Macharet2016}, or on quantum Overhauser magnetometers \cite{Sapunov2025}. However, the cutting edge compact AMs enable one- to three-order-of-magnitude improvements in both sensitivity and bandwidth of measurements (a comparative review can be found in Ref.~\cite{Zheng2021}).

%They have found numerous applications both in fundamental physics (e.g., the search for dark matter \cite{Afach2021}) and in solving important applied problems, ranging from medical diagnostics \cite{Schofield2025,Fedosov2025} to archeology \cite{Schmidt2024} and non-destructive materials inspection \cite{Bevington2018}.

%For example, the GEM GSMP-35U potassium AM features a sensitivity of $300$~fT/$\surd$Hz and a bandwidth of 10 Hz, with a sensor weight of only 430 g.

In OPMs used for geophysical applications, various techniques are employed to excite magnetic resonance (MR) in alkali-metal atoms (Rb, Cs, or K). The most common types of commercial quantum magnetometers include optically pumped radio-frequency (rf) magnetometers of $M_x$ and $M_z$ types (see review \cite{Alexandrov2013}). For airborne magnetic surveys, $M_x$ OPMs are generally preferred, as their response bandwidth exceeds that of $M_z$ ones. However, there is currently growing interest in all-optical OPMs, as their performance characteristics are not only comparable to those demonstrated by rf AMs but can also substantially surpass them when the magnetometer is miniaturized. For instance, commercially available highly compact OPMs based on the all-optical free-induction-decay (FID) technique exhibit sensitivities of a few pT/$\surd$Hz and bandwidths of up to $500$~Hz \cite{QuSpin}. In a gradiometric configuration, when two cells are used, the sensitivity can be improved to $0.2$~pT/$\surd$Hz \cite{Twinleaf}.

Another all-optical technique, known as the Bell--Bloom scheme, yields comparable results. For example, in Ref.~\cite{Gerginov2017}, using a miniature vapor cell, a sensitivity of $1$~pT/$\surd$Hz was achieved in a single-channel configuration and $0.2$~pT/$\surd$Hz in a gradiometric configuration, with a measurement bandwidth of $750$~Hz. A comprehensive comparison of the rf $M_x$ and all-optical techniques is presented in Ref.~\cite{Schultze2012}.

%At the same time, $M_z$ AMs are simpler to implement and less susceptible to phase noise, therefore, they are also used in magnetic surveys \cite{Zhou2026}.

%Another type of an all-optical sensor based on the Bell-Bloom technique has recently shown breakthrough results in navigation \cite{Muradoglu2025}.

All-optical OPMs offer several advantages for a wide range of UAV-borne applications. Such a magnetometer does not require an rf field, thereby eliminating any rf interference. This is particularly beneficial when multiple closely spaced sensors are used. The absence of rf coils enables a more compact sensor head, reducing power consumption. Furthermore, all-optical OPMs exhibit just a single (``polar'') dead zone, rather than two dead zones typically present in rf $M_x$ OPMs \cite{Schultze2012}. This allows UAVs to maneuver more freely without signal loss, thereby improving the quality of the acquired data. Finally, single-beam all-optical schemes, which are particularly well suited for UAV-borne applications, have a simpler design than their single-beam $M_x$-magnetometer counterparts. Therefore, further development of all-optical quantum technology for high-sensitivity magnetic field measurements is of considerable practical interest.

%All-optical magnetometry is likewise of significant interest for medical diagnostics, where Hanle-effect magnetic sensors operating in the spin-exchange relaxation-free (SERF) regime are commonly employed. These sensors, however, are restricted to near-zero fields, suffer from a low dynamic range, require relatively high cell temperatures ($150^\circ$C or higher), and demand careful crosstalk suppression between neighboring sensor heads. All-optical sensors could circumvent many of these limitations. Although no commercial systems currently exist, research activities in this area remain substantial.

%For example, a sensor head of the all-optical QTFM Gen-2 magnetometer by QuSpin \cite{QuSpin} features extremely miniature volume of $\approx\,$$18$$\times$$20$$\times$$35$~mm$^3$.

%On the contrary, in some cases, all-optical AMs have already surpassed rf AMs.

%However, a breakthrough result can be attributed to the recent work \cite{Muradoglu2025}, in which a compact magnetometer sensor head ($150$~cm$^3$) was demonstrated, implementing the Bell–Bloom method with amplitude-modulated light and orthogonal probe and pump light beams. Under field conditions, the magnetometer exhibited a remarkable sensitivity of $0.08$~pT/$\surd$Hz and a dynamic range of $250$~Hz. 

In this work, a novel Bell--Bloom-like approach for high-sensitivity measurements is proposed and studied. A single elliptically polarized light beam is used both for pumping and probing atoms confined in a $0.125$~cm$^3$ glass vapor cell. The beam is tuned to resonance with the cesium D$_1$ line. A similar configuration was studied in Refs.~\cite{Shah2009,Tang2021}; however, the authors proposed a magnetic sensor scheme based on the ground-state Hanle effect, operating near zero magnetic field under the spin-exchange relaxation-free (SERF) regime \cite{Happer1977}. This regime is primarily intended for medical diagnostics in environments requiring a high degree of magnetic shielding, where the residual magnetic field must be below $1$~nT \cite{Borna2022}. Another compact OPM based on the Bell--Bloom technique with a single elliptically polarized light beam has recently been considered in Ref.~\cite{Lee2026} for magnetic sensing in geomagnetic-field environments. The authors demonstrated a sensitivity of $6$~pT/$\surd$Hz at a $200$~Hz bandwidth. As in previous works \cite{Shah2009,Tang2021}, the laser frequency was detuned far from the center of the absorption line, so that magnetic resonance (MR) was induced by the circular birefringence of the medium, leading to rotation of the probe wave's linear polarization. Such configurations may introduce spurious ellipticity in a linearly polarized probe wave, resulting in errors in magnetic field measurements \cite{Li2017}. Furthermore, stabilizing the optical frequency far from the absorption-line center is more challenging than stabilizing it near the line center, requiring more technically complex solutions \cite{Hu2017}. Such stabilization is essential for minimizing the detrimental effects associated with optical-frequency drift.

%Such an approach suffers from several adverse effects, including optical frequency drift and increased sensitivity to radiation phase noise. 

%A magnetic resonance was observed in the rotation of the light polarization ellipse caused by circular birefringence of the medium. In our case, a magnetic resonance is associated with circular dichroism of the medium and is observed in an ellipticity parameter of the light wave polarization.

In another study \cite{Sun2017}, an intriguing Bell--Bloom-like scheme was proposed, enabling vector (3D) magnetic field measurements. In that work, magnetic resonance was associated with the circular dichroism of the medium and was observed as a change in the ellipticity parameter of light. The measurement sensitivity was estimated at $80$~fT/$\surd$Hz. However, achieving this notably high sensitivity necessitated the use of a relatively large spherical Cs vapor cell, $30$~mm in diameter, as well as a configuration involving not a single light beam but two linearly polarized probe beams and one circularly polarized pump beam propagating in mutually orthogonal directions. Such a scheme is unsuitable for the implementation of a miniaturized sensor.

Here, we study an alternative Bell--Bloom-like technique that shares a few features with that investigated in our previous work \cite{Makarov2025}. In this approach, the light field is in resonance with the medium, while MR is induced by circular dichroism rather than birefringence. In Ref.~\cite{Makarov2025}, an amplitude-modulated light beam produced with the help of an acousto-optic modulator (AOM) was employed to observe MR in rubidium atoms. The light beam was generated by an external-cavity diode laser (ECDL). However, that scheme is unsuitable for miniaturization, consumes excessive power, and is highly sensitive to external perturbations.

In the present work, we propose an improved and simplified scheme that can be both highly compact and robust. Instead of relying on bulky components such as an AOM and an ECDL, we employ a single vertical-cavity surface-emitting laser (VCSEL)~\cite{Blokhin2019}. Modulation of the injection laser current results in frequency modulation (FM) of the laser emission, which is used to excite and observe a MR. Moreover, in contrast to our previous work~\cite{Makarov2025}, the theoretical model presented here explicitly accounts for the transformation of the light polarization in the polarimeter, as well as for an arbitrary orientation of the magnetic-field vector with respect to the laser-beam wave vector. The latter allows us to characterize the sensor’s dead zone.

% Numerous other Bell--Bloom-like schemes with FM of the light field have already proven highly effective in magnetometry \cite{Budker2002,Schwindt2005,Acosta2006,Balabas2006,Martinez2010,Lee2026}. 
%It should be noted that a similar principle of the sensor operation was studied by Sebbag et al. \cite{Sebbag2021}. However, the authors examined a vector magnetic sensor based on the ground-state Hanle effect under the SERF regime of operation, which was aimed to perform measurements near zero magnetic field. In our work, we study a scalar sensor scheme, performing measurements of the Earth’s scale magnetic fields. Moreover, we use an additional optical element, a half-wave plate, which helps to equalise the probe and pump wave transmission signals, improving SNR of the sensor.

%It is worth noting that numerous other FM Bell--Bloom--like schemes have already proven highly effective in magnetometry

The paper is organized as follows. In Section~\ref{sec:2}, we provide a concise theoretical description of the proposed technique. Next, we verify the proposed technique experimentally in proof-of-concept experiments presented in Section~\ref{sec:3}. In particular, under the current experimental conditions, the sensitivity is estimated to be approximately $220$~fT/$\sqrt{\mathrm{Hz}}$ in an ambient magnetic field of $\approx\,$$50$~$\mu$T, while the bandwidth ($-3$~dB) is approximately $1$~kHz. The sensor has an angular operating range of approximately $\pm$$35^\circ$ about the optimal magnetic-field orientation (${\bf B}$$\perp$${\bf k}$).

\section{\label{sec:2}Theoretical background}

\subsection{\label{sec:21}Configuration of the fields}

%For compact devices, single-beam configurations based on buffer-gas-filled small ($V$$\,<\,$$1$~cm^3) vapor cells are preferred, as they offer greater reliability and are simpler to implement \cite{Shah2009,Tang2021,Makarov2025,Petrenko2021,Sebbag2021,Li2021,Petrenko2022,Petrenko20222,Lee2026}.

\begin{figure}[!t]
\centering
\includegraphics[width=0.6\linewidth]{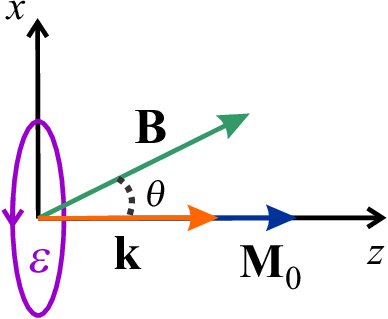}
\caption{\label{fig:1} Geometry of the fields. The light wave travels along the $z$-axis, while the magnetic field ${\bf B}$ lies in the $(x,z)$ plane. Here, ${\bf k}$ is the wave vector, $\epsilon$ is the ellipticity of the light polarization, $\theta$ is the angle between the vectors ${\bf B}$ and ${\bf k}$, ${\bf M}_0$ is the macroscopic magnetization of the vapor produced by the light field in the absence of the magnetic field and relaxation processes}
\end{figure}

A polarimetric technique for detecting MRs is widely used in magnetic sensors to suppress laser intensity noise, thereby enhancing the signal-to-noise ratio (SNR) and improving measurement sensitivity \cite{Budker2007}. For compact devices, single-beam configurations based on buffer-gas-filled small ($V$$\,<\,$$1$~cm$^3$) vapor cells are preferred, as they offer greater reliability and are simpler to implement \cite{Shah2009,Tang2021,Lee2026,Makarov2025,Petrenko2021,Sebbag2021,Petrenko2022,Petrenko20222}. Most such sensors rely on the effect of nonlinear magneto-optical rotation (NMOR) of an elliptically polarized light beam, which arises from the magnetically sensitive circular birefringence of the medium \cite{Auzinsh2010}. In the present work, however, we employ magnetically sensitive circular dichroism, which causes a change in the ellipticity of the light polarization.

It is natural to represent an elliptically polarized beam as a superposition of two waves with opposite circular polarizations propagating along the $z$‑axis (see Fig.~\ref{fig:1}):

%The elliptically polarised beam can be decomposed into two components: the circularly polarised component optically pumps the atoms into the end Zeeman (``stretched'') state, whereas the linearly polarised component serves as a probe wave and undergoes Faraday rotation.

\begin{equation}\label{lightfield}
    {\bf E}(z,t) = \Bigl(E_c(t)\,\,{\bf e}_{+1} + E_p(t)\,{\bf e}_{-1}\Bigr)\,e^{-i(\omega t - k z)} + c.c.,
\end{equation}

\noindent where ``\textit{c.c.}'' means a complex conjugate term, $\omega$ is the optical angular frequency, $k$$\,=\,$$\omega/c$ is the magnitude of the wave vector, with $c$ being the speed of light. $E_c$ and $E_p$ are the slowly varying amplitudes of the pump (coupling) and probe waves, respectively. The complex basic spherical vectors ${\bf e}_{\pm1}$ are related to the Cartesian ones (${\bf e}_{x,y}$) as follows \cite{Varshalovich1988}: ${\bf e}_{\pm1} = \mp({\bf e}_x \pm i {\bf e}_y)/\sqrt{2}$. 

In Eq. (\ref{lightfield}), we assume that the more intense wave has right-hand circular polarization, whereas the probe wave has left-hand circular polarization. The former induces $\sigma^+$ dipole transitions in the atom, whereas the latter associates with $\sigma^-$ dipole transitions, while only weakly perturbing the atomic quantum state in the presence of the stronger pump wave. The amplitudes of the pump and probe waves are $E_c$$\,=\,$$E_0(t)\,$$\cos(\epsilon$$-$$\pi/4)$ and $E_p$$\,=\,$$E_0(t)\,$$\sin(\epsilon$$-$$\pi/4)$, respectively, with $\epsilon$ being the polarization ellipticity parameter. It can take values ranging from $-\pi$$/4$ to $\pi$$/4$, where $\pm\pi$$/4$ correspond to left- and right-circularly polarized light, respectively, and $\epsilon$$\,=\,$$0$ corresponds to linear polarization. The absolute value of $E_0$ determines the total intensity of the light wave, $I_0$$\,=\,$$(c/2\pi)$$\,|E_0|^2$, while the pump and probe wave intensities are $I_{c,p}$$\,=\,$$(c/2\pi)$$\,|E_{c,p}|^2$. Since the beam diameter ($d$) is fixed in our experiments, we likewise use the optical power as a measure of the optical intensity according to the relation: $P_0$$\,=\,$$\pi$$d^2$$I_0$$/4$.

\begin{figure}[!t]
\centering
\includegraphics[width=1\linewidth]{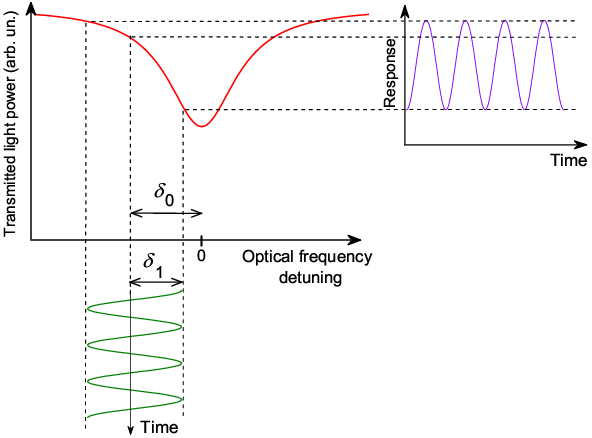}
\caption{\label{fig:2} Qualitative illustration of the FM Bell--Bloom technique. Here, $\delta_0$ is the optical frequency detuning from the center of the absorption line, and $\delta_1$ is the amplitude of frequency modulation (frequency deviation). Further details are given in the text}
\end{figure}

The external magnetic field, ${\bf B}$, is assumed to be directed arbitrarily. However, we can choose Cartesian coordinates in such a way that the magnetic field vector lies in the $xz$-plane (see Fig.~\ref{fig:1}). Therefore, it can be represented as follows: ${\bf B}$$\,=\,$$B_x$${\bf e}_x$$+$$B_z$${\bf e}_z$. The Cartesian components of the ${\bf B}$ field, in turn, can be written in the following form: $B_x$$\,=\,$$B_0$$\,\sin(\theta)$, $B_z$$\,=\,$$B_0$$\,\cos(\theta)$ with $\theta$ being the angle between the magnetic field vector and the $z$-axis, $B_0$ being the absolute value of the magnetic field vector.

The Bell--Bloom method has been extensively investigated using amplitude \cite{Bell1961,Balabas2006,Pustelny2008,Li2022}, frequency \cite{Budker2002,Acosta2006,Balabas2006,Martinez2010,Lee2026}, or polarization \cite{Gilles1991,Breschi2014,Bevilacqua2014,Petrenko2021,Petrenko2022,Petrenko20222,Pustelny2025} modulation of the optical field. For compact quantum sensors, however, FM Bell--Bloom schemes are particularly advantageous, as they can be implemented solely through modulation of the diode-laser pump current, thus eliminating the need for external modulators. Accordingly, in the present work we employ this type of modulation, which is schematically illustrated in Fig.~\ref{fig:2}. Here, we use sinusoidal FM, although other forms of the modulating signal, such as a square wave \cite{Martinez2010,Gerginov2017}, can also be employed. As shown in the figure, for an appropriate choice of the optical frequency detuning ($\delta_0$) and the amplitude of modulation ($\delta_1$), the response of the medium detected by a photodetector exhibits a sinusoidal-like modulation as well. In our experiments, we use precisely this FM regime.

\begin{figure}[!b]
\centering
\includegraphics[width=1\linewidth]{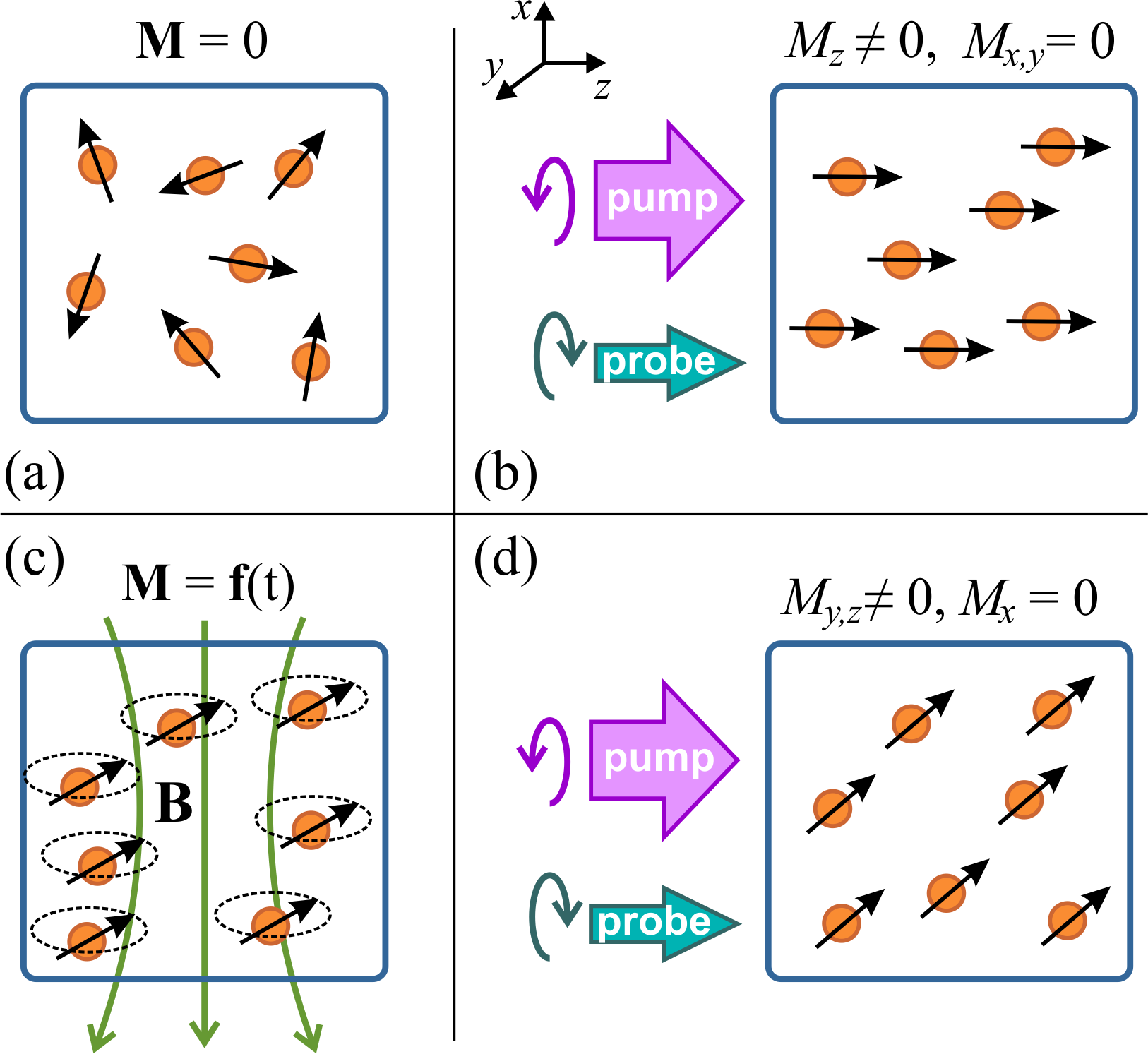}
\caption{\label{fig:3} Qualitative illustration of the FM Bell--Bloom technique using an elliptically polarized light beam. (a) In the absence of pump light, the atomic spins are randomly oriented in space. (b) A circularly polarized pump wave generates macroscopic magnetization (atomic spin orientation) along ${\bf k}$. (c) When the light is switched off, the magnetization vector freely precesses about the external magnetic field. (d) When the light field is switched on again, it interacts with the atoms whose spins are oriented at a certain angle with respect to ${\bf k}$}
\end{figure}

\subsection{\label{sec:22}Qualitative explanation of the resonance}

We first describe the physics underlying the formation of the magnetic resonance in our scheme based on Fig.~\ref{fig:3}. In the absence of light, the medium is isotropic, meaning that the average magnetic dipole moment, ${\bf M}$, is zero [Fig.~\ref{fig:3}(a)]. When exposed to a resonant circularly polarized pump wave, the atomic spins become oriented along the $z$-axis, generating a macroscopic magnetic dipole moment [Fig.~\ref{fig:3}(b)]. In this case, the absorption of the pump wave is reduced.

When the optical field is far detuned from resonance with the medium, the atoms interact negligibly with the light, and therefore the vector ${\bf M}$ undergoes free (Larmor) precession about the magnetic-field vector ${\bf B}$ [Fig.~\ref{fig:3}(c)]. In the figure, this field is directed along the $x$-axis, corresponding to the configuration in which the sensor exhibits its highest sensitivity (see the operational zone measurements in Section~\ref{sec:3}). The precession frequency is determined by the magnitude of ${\bf B}$ field as $\Omega_L$$\,=\,$$\gamma$$B_0$ with $\gamma$ being the gyromagnetic ratio.

%If the atomic spins are oriented along this vector, the pump wave, as in panel (b), is weakly absorbed by the medium.

When the pump wave is tuned back into resonance with the atoms, it is absorbed by the medium, with the absorption strength depending on the instantaneous orientation of ${\bf M}$ [Fig.~\ref{fig:3}(d)]. Importantly, if ${\bf M}$ completes a full Larmor precession cycle while the pump wave remains far detuned from resonance, the atoms will again be oriented along the $z$-axis upon returning to resonance, as in Fig.~\ref{fig:3}(b), and the pump-wave absorption will again be reduced. Consequently, as the FM frequency $f_m$ is scanned near $\Omega_L$, the amplitude of the sinusoidal response signal in Fig.~\ref{fig:2} (violet curve) reaches a maximum at $f_m$$\,=\,$$\Omega_L$. Thus, an MR is formed in the amplitude of the sinusoidal response of the medium. Using standard synchronous detection, the amplitude of the sinusoidal response can be measured separately, allowing the modulation frequency $f_m$ corresponding to the center of the magnetic resonance, $f_m$$,=,$$\Omega_L$, to be determined. The latter directly provides the magnitude of the external magnetic field.

It is evident that the behavior of the probe-wave absorption in the medium when scanning $f_m$ is opposite to that of the pump-wave absorption; i.e., at $f_m$$\,=\,$$\Omega_L$ the probe wave experiences maximum absorption (see also the discussion in Ref.~\cite{Makarov2025}). By employing a polarimeter, one can separately monitor the transmitted pump and probe wave intensities to observe MRs of opposite signs. Then, using a balanced photodetector, these oppositely signed resonances enable suppression of laser intensity noise, since at the balanced output of the photodetector, this noise is significantly reduced compared to that in each individual channel of the photodetector. A similar effect of intensity noise suppression occurs in OPMs based on nonlinear magneto-optical rotation (NMOR) \cite{Li2008}.

%However, in our case, there is no need to employ a light field far detuned from the center of the absorption line.

%(a) Atoms in the absence of the light field. The macroscopic magnetic moment of the medium is zero. (b) The pump field polarizes the atomic spins along the $z$-axis. (c) During the interval, when the optical field is far from resonance with the medium, the magnetic moment undergoes Larmor precession in the external magnetic field. Here, the magnetic field is directed along the $x$-axis, corresponding to the configuration in which the sensor exhibits its highest sensitivity. (d) The pump and probe waves experience different strengths of absorption in the medium depending on the orientation of the magnetic moment.

\subsection{\label{sec:23}Bloch equation formalism}

For our theoretical analysis, we assume the probe-wave intensity to be sufficiently low that its influence on the atomic state can be neglected, corresponding to the linear interaction regime for the probe field. Being circularly polarized, the pump wave polarizes the atomic spins along the $z$-axis, thereby inducing a macroscopic dipole magnetization ${\bf M}$ in the medium, also known as the first-rank polarization moment (orientation). In atomic magnetometry, a description of the magnetization vector behavior in an ambient magnetic field is often provided by an intuitive and simplified formalism based on the Bloch equation of motion \cite{Dehmelt1957,Bell1961}:

\begin{equation}\label{BlochEq}
    \frac{d{\bf M}}{dt} = \gamma \bigl[{\bf M}\times{\bf B}\bigr] + \mathcal{R}(t)\,\bigl({\bf M}_0-{\bf M}\bigr) - \Gamma\, {\bf M}\,.
\end{equation}

\noindent Here, $\gamma$$\,\approx\,$$2\pi$$\times$$3.5$~Hz/nT for the cesium atom, while $R$ denotes the optical pumping rate of the atomic ground state induced by the pump wave ($\mathcal{R}$$\,\propto\,$$I_c$). For the sake of simplicity, we assume that the transverse and longitudinal relaxation rates are equal, and thus a single relaxation constant $\Gamma$ is introduced to describe these processes. In the steady-state regime, in the absence of the magnetic field and relaxation, the wave ${\bf E}_c$ induces a magnetic moment ${\bf M}_0$ along the $z$-axis, i.e., ${\bf M}_0$$\,=\,$$\{0,0,\mathcal{M}\}^{\rm T}$ (see Fig.~\ref{fig:1}).

The FM regime of excitation shown in Fig.~\ref{fig:2} produces a medium response similar to that observed under amplitude modulation of light tuned to the center of the absorption line~\cite{Makarov2025}. To demonstrate this mathematically, consider the expression for the light-wave transmission coefficient ($\chi$) in a buffer-gas-filled vapor cell, governed by a Lorentzian absorption profile:

\begin{equation}\label{eq:Tcoeff}
    \chi = 1 - \alpha(\delta) = 1 - \frac{\alpha_0\Gamma_0^2}{\Gamma_0^2+\delta^2}\,,
\end{equation}

\noindent where $\alpha$ is the absorption coefficient and $\alpha_0$ is its value at resonance ($\delta$$\,=\,$$0$), which depends on the temperature of alkali-metal atoms and the buffer-gas pressure. $\Gamma_0$ is the half-width at half maximum (HWHM) of the absorption line, which is determined mainly by the buffer-gas collisional broadening. The transmission coefficient reaches its minimum value of $1$$-$$\alpha_0$ at exact resonance and remains unity away from resonance.

We next expand Eq.~(\ref{eq:Tcoeff}) in a Taylor series, retaining only the term linear in $\delta$, since in the experiments the optical frequency detuning corresponds to the slope of the resonance (see Fig.~\ref{fig:2}):

\begin{equation}\label{eq:TcoeffApprox}
    \chi \approx 1 + \alpha_0\Big(\frac{\delta}{2\Gamma_0}-1\Big)\,.
\end{equation}

\noindent In the case of FM, the detuning of the laser frequency from the center of the absorption line varies harmonically in time:

\begin{equation}\label{eq:detunmodul}
    \delta(t) = \delta_0 + \delta_1 \cos(f_m t)\,.
\end{equation}

\noindent Substituting (\ref{eq:detunmodul}) into (\ref{eq:TcoeffApprox}) and taking into account that the conditions $\delta_0$$\,\approx\,$$-\delta_1$$\,\approx\,$$-\Gamma_0$ are satisfied in our experiments, we arrive at the final expression:

\begin{equation}\label{eq:TcoeffApprox2}
    \chi \approx 1 - \frac{\alpha_0}{2} - \frac{\alpha_0}{2}\cos(f_m t)\,.
\end{equation}

\noindent Thus, at the initial moment, $f_m t$$\,=\,$$0$, the medium is least transparent ($\chi$$\,=\,$$1$$-$$\alpha_0$), whereas at $f_m t$$\,=\,$$\pi$, the radiation is transmitted through the medium without attenuation ($\chi$$\,\approx\,$$1$). Eq. (\ref{eq:TcoeffApprox2}) suggests that the optical pumping rate in (\ref{BlochEq}) can be modeled in the following simple way:

\begin{equation}\label{eq:RateModul}
    \mathcal{R}(t) = \mathcal{R}_0 \big[1 + \cos(f_m t)\big]\,.
\end{equation}

\noindent This behavior is equivalent to that employed in Bell--Bloom schemes with amplitude modulation of the light~\cite{Bell1961,Makarov2025}.

To obtain an analytical solution of Eq. (\ref{BlochEq}), it is convenient to rewrite Eq. (\ref{eq:RateModul}) in the form:

\begin{equation}\label{harmonicsR}
    \mathcal{R}(t) = \sum_{n=-\infty}^{\infty}\mathcal{R}^{(n)} e^{inf_mt}\,,
\end{equation}

\noindent where only the zeroth- and first-order harmonics are nonzero:

\begin{equation}\label{harmonicsR2}
    \mathcal{R}^{(n)} = \mathcal{R}_0\,\delta_{n,0} +\frac{\mathcal{R}_0}{2}\big[\delta_{n,+1} + \delta_{n,-1}\big]\,,
\end{equation}

\noindent with $\delta_{n,m}$ is the Kronecker delta.

The harmonics in (\ref{harmonicsR}) result in the appearance of the corresponding harmonics in the magnetization vector:

\begin{equation}\label{harmonicsM}
    {\bf M}(t)=\sum_{n=-\infty}^{\infty}{\bf M}^{(n)} e^{inf_mt}\,,
\end{equation}

\noindent with ${\bf M}^{(n)*}$$\,=\,$${\bf M}^{(-n)}$. The nearly sinusoidal modulation of the response in Fig.~\ref{fig:2} suggests that, for our qualitative analysis, it is sufficient to retain only the the lowest-order harmonics $M^{(n)}$ with $n$$\,=\,$$0,$$\,\pm1$.

The Bloch equation can be written in the matrix form:

\begin{eqnarray}\label{BlochEq2}
    inf_m{\bf M}^{(n)} &=& \mathcal{R}^{(n)}{\bf M}_0 - \hat{\mathcal{L}}\,{\bf M}^{(n)}\nonumber\\
    && -\Gamma{\bf M}^{(n)}-\sum_{q=-1}^{+1}R^{(q)}{\bf M}^{(n-q)}\,,
\end{eqnarray}

\noindent Here, the antisymmetric matrix $\hat{\mathcal{L}}$ governs the Larmor precession and is given by:

\begin{equation}
    \hat{\mathcal{L}}=\left(
\begin{array}{ccc}
0 & -\Omega_z & 0 \\
\Omega_z & 0 & -\Omega_x \\
0 & \Omega_x & 0
\end{array}\right)\,,
\end{equation}

\noindent where $\Omega_{x,z}$$\,=\,$$\gamma B_{x,z}$. We solved Eq.~(\ref{BlochEq2}) in the low-order harmonic approximation, retaining only the zeroth ($n$$\,=\,$$0$) and first ($n$$\,=\,$$\pm1$) harmonics. The details of the routine calculations will be published elsewhere. 

Let us analyze the intensities of the pump and probe waves after the vapor cell, $I_c$ and $I_p$, as the functions of $f_m$. It is known that the light wave absorption in the vapor cell depends on the $z$-component of the ${\bf M}$ vector as follows \cite{Bell1961,Shah2009}: $1$$\,\mp\,$$M_z/\mathcal{M}$, where ``$-$'' stands for the pump wave, whereas ``$+$'' is responsible for the probe one. Here, we have neglected the contribution of the quadrupole polarization moment in the atomic ground state, which is a common approximation in the low-power limit~\cite{Castagna2011}. In fact, the sign at $M_z$ is determined not only by the polarization of the light wave but also by the type of the atomic transition (see discussion in Ref.~\cite{Brazhnikov2026}). In particular, for ``bright'' and ``dark'' transitions the sign may differ. A quantitative analysis of these features, however, requires a more rigorous treatment employing the density matrix formalism, which is beyond the scope of the present analysis. Therefore, we can consider the following commonly used expression \cite{Makarov2025}:

%The modulation of the light intensity in the cell arising from the modulation of the frequency detuning should be also taken into account.

%Let us analyze the absorption profiles of the pump and probe waves, $W_c$ and $W_p$, respectively. Both signals depend on the $z$-component of the ${\bf M}$ vector as follows \cite{Shah2009}:

%\begin{equation}\label{signal}
%W_{c,p}(t) \propto I_{c,p}\Bigl[1 \mp %M_z(t)/\mathcal{M}\Bigr]\,,
%\end{equation}

\begin{eqnarray}\label{signal}
I_{c,p}(t) \approx I_{c0,p0}\Bigl\{1-\alpha(t)\ell\Bigl[1 \mp M_z(t)/\mathcal{M}\Bigr]\Bigr\}\,,
\end{eqnarray}

%Hereafter, for simplicity, we set $\mathcal{M}$$\,=\,$$1$. 

\noindent where $I_{c0,p0}$ are the wave intensities at the cell entrance, $\ell$ is the cell length. Similarly to Eq. (\ref{eq:TcoeffApprox2}), the absorption coefficient $\alpha(t)$ can be expressed as follows:

\begin{equation}\label{alpha}
\alpha(t) \approx \frac{\alpha_0}{2}\big[1+\cos(f_mt)\bigr]\,.
\end{equation}

In practice, MR is observed employing a synchronous modulation/demodulation technique with the help of a lock-in amplifier (LIA). Most commonly, when the pump wave is circularly polarized, the signal at the first harmonic, i.e., at the frequency $f_m$, is analyzed:

\begin{equation}
I(t) = I^{(in)} \cos(f_mt) - I^{(out)} \sin(f_mt)\,,
\end{equation}

\noindent where $I^{(in)}$ and $I^{(out)}$ are the in-phase and out-of-phase (quadrature) components of the signal with respect to the frequency detuning modulation (\ref{eq:detunmodul}). In lock-in amplifiers, these signals are commonly referred to as the $X$ output and $Y$ output, respectively. Based on Eqs. (\ref{harmonicsM}), (\ref{signal}) and (\ref{alpha}), the in-phase and quadrature components of the pump and probe intensities are given by

\begin{figure*}[!t]
\centering
\includegraphics[width=1\linewidth]{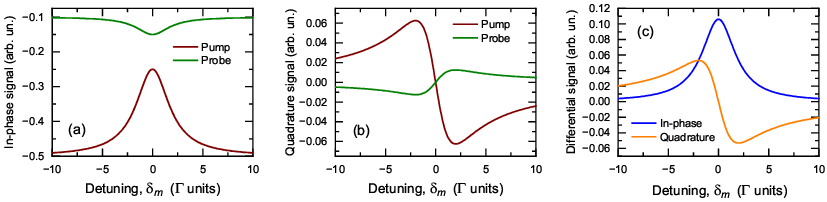}
\caption{\label{fig:4} Resonance curves calculated according to Eqs. (\ref{Winc})--(\ref{Woutp}) for the pump and probe waves monitored separately: (a) in-phase components, (b) out-of-phase components. Panel (c) shows the in-phase and out-of-phase components of the signal from the differential (balanced) photodetector output calculated according to Eqs. (\ref{eq:Iindiff}), (\ref{eq:Ioutdiff}). Parameters for modeling: $\alpha_0 L_{\rm cell}$$\,=\,$$1$, $I_{c0}$$\,=\,$$1$, $I_{p0}$$\,=\,$$0.2$, $\xi$$\,=\,$$1$, $\theta$$\,=\,$$\pi/2$, $\eta_1$$\,=\,$$\pi/16$}
\end{figure*}

\begin{eqnarray}
&&I^{(in)}_{c,p} = -\frac{\alpha_0 \ell I_{c0,p0}}{2}\,\Biggl[1 \mp \frac{M_z^{(0)}}{\mathcal{M}} \mp \frac{M_{1z}}{\mathcal{M}}\Biggr]\,,\label{Win}\\
&&I^{(out)}_{c,p} = \mp \frac{\alpha_0 \ell I_{c0,p0}}{2}\,\frac{M_{2z}}{\mathcal{M}}\,.\label{Wout}
\end{eqnarray}

\noindent Here, the upper sign corresponds to the pump wave ($I_c$), whereas the lower sign pertains to the probe one ($I_p$). $M_{1z}$ and $M_{2z}$ are the $z$-components of the vectors ${\bf M}_1$ and ${\bf M}_2$ defined as the following real and imaginary parts:

\begin{equation}\label{eq:M1M2}
    {\bf M}_1 = 2 {\rm Re} \Big\{{\bf M}^{(+1)}\Big\},\,\, {\bf M}_2  = -2 {\rm Im} \Big\{{\bf M}^{(+1)}\Big\}.
\end{equation}

Based on the solution derived from Eq.~(\ref{BlochEq2}), we arrive at the final expressions for the in-phase and quadrature components of the pump- and probe-wave intensities after the medium:

\begin{eqnarray}
&&I^{(in)}_c = -\frac{\alpha_0 \ell I_{c0}}{2}\Biggl[1-\frac{I_{c0}\xi\cos^2(\theta)}{\Delta}\nonumber\\
&&\qquad\qquad\qquad\qquad\qquad-\frac{I_{c0}\xi\Delta\sin^2(\theta)}{2\big(\delta_m^2+\Delta^2\big)}\Biggr]\,,\label{Winc}\\[4pt]
&&I^{(in)}_p = -\frac{\alpha_0 \ell I_{p0}}{2}\Biggl[1+\frac{I_{c0}\xi\cos^2(\theta)}{\Delta}\nonumber\\
&&\qquad\qquad\qquad\qquad\qquad+\frac{I_{c0}\xi\Delta\sin^2(\theta)}{2\big(\delta_m^2+\Delta^2\big)}\Biggr]\,,\label{Winp}\\[4pt]
&&I^{(out)}_c = -\frac{\alpha_0 \ell I_{c0}^2\,\xi\delta_m\sin^2(\theta)}{4\bigl(\delta_m^2+\Delta^2\bigr)}\,,\label{Woutc}\\[4pt]
&&I^{(out)}_p = \frac{\alpha_0 \ell I_{p0}I_{c0}\,\xi\delta_m\sin^2(\theta)}{4\bigl(\delta_m^2+\Delta^2\bigr)}\,,\label{Woutp}
\end{eqnarray}
 
\noindent where $\Delta$$\,=\,$$\mathcal{R}_0$$\,+\,$$\Gamma$ is the HWHM of the magnetic resonance, $\delta_m$$\,=\,$$f_m$$\,-\,$$\Omega_L$ is the detuning of the modulation frequency from the Larmor frequency. The coefficient $\xi$ depends on the optical transition strength in the D$_1$ line and is determined by the following expression: $\mathcal{R}_0$$\,=\,$$\xi$$I_{c0}$. Eqs.~(\ref{Winc})--(\ref{Woutp}) are valid under the assumption $f_m$$\sim$$\Omega_L$$\,\gg\,$$\mathcal{R}_0,$$\,\Gamma$, which is consistent with the typical experimental conditions under the Earth's magnetic field ($\approx\,$$50$~$\mu$T).

It follows from Eqs.~(\ref{Winc})--(\ref{Woutp}) that MR observed in the in-phase channel of the LIA is described by a Lorentzian absorptive line shape, whereas MR in the quadrature channel is characterized by a Lorentzian dispersive profile. These profiles are modeled in Fig.~\ref{fig:4}. The in-phase signals, Fig.~\ref{fig:4}(a), contain a finite constant background, whereas the quadrature signals, Fig.~\ref{fig:4}(b), have no such background. Physically, the negative value of the background in the in-phase signals arises from the fact that the harmonic response of the medium and the optical detuning modulation described by Eq.~(\ref{eq:detunmodul}) have a relative phase difference of $\pi$, as follows from Fig.~\ref{fig:2}.

Either of the plotted signals in Fig.~\ref{fig:4} can be used in the magnetic sensor for measurements. Nevertheless, the quadrature channel is most naturally adopted, as it inherently yields an error signal suitable for automatic locking of the generator frequency ($f_m$) to the Larmor frequency ($\Omega_L$) and for the subsequent evaluation of the absolute value of magnetic field. Obviously, the sensitivity of magnetic-field measurements increases with increasing resonance height and decreasing resonance linewidth. Thus, the height-to-linewidth ratio plays a crucial role in achieving high sensor sensitivity. As an example, consider the MR height in the in-phase channel (\ref{Winc}):

\begin{equation}\label{eq:AmpPump}
    A = \frac{\alpha_0 \ell I_{c0}^2\,\xi \sin^2(\theta)}{4\Delta}\,.
\end{equation}

\noindent It follows that the height depends quadratically on the intensity and reaches its maximum at $\theta$$\,=\,$$\pi/2$, corresponding to the $\mathbf{B}$$\perp$$\mathbf{k}$ configuration.

If the quadrature channel of the LIA is used for magnetic-field measurements, the sensitivity increases with the slope of the dispersive-like MR resonance in $I^{(out)}_c$ at $\delta_m$$\,=\,$$0$. As can be readily shown, this slope is given by

\begin{equation}\label{eq:Slope}
    S = \frac{\alpha_0 \ell I_{c0}^2\,\xi \sin^2(\theta)}{4\Delta^2} = \frac{A}{\Delta}\,.
\end{equation}

\noindent Thus, the slope of the MR in the quadrature channel is just equal to the ratio of the MR height to its half-width in the in-phase channel.

Therefore, the sensitivity of the magnetic sensor, defined as the smallest change in the magnetic field that can be detected within a $1$-Hz bandwidth, can be estimated from the following equivalent expressions~\cite{Martinez2010}:

\begin{equation}\label{eq:Sensitivity}
    \delta B \approx \frac{\mathcal{N}}{S} = \frac{\mathcal{N}}{(A/\Delta)} = \frac{\Delta}{\rm SNR}\,,
\end{equation}

\noindent where $\mathcal{N}$ is the noise voltage spectral density, and SNR is the signal-to-noise ratio within a $1$-Hz bandwidth, defined as $A$$/$$\mathcal{N}$.

To utilize, for instance, the quadrature signal $I_c^{(out)}$ for magnetic field measurements, it is necessary to separate it from the signal $I_p^{(out)}$. Since the probe wave and the pump wave possess opposite circular polarizations, this separation can be accomplished relatively simply using a polarimeter, as already mentioned in Sec.~\ref{sec:22}. The polarimeter consists of a quarter-wave ($\lambda/4$) plate, a Wollaston prism (WP), and a balanced photodetector (BPD). The $\lambda/4$ plate converts each of the circularly polarized waves into a linearly polarized wave, with the resulting polarizations being mutually orthogonal. The Wollaston prism then directs each of the two linearly polarized waves to a separate photodetector for subsequent use in magnetic field measurements.

In practice, however, improving the measurement sensitivity often requires increasing the SNR by suppressing so-called common-mode noise. The dominant contribution to this noise typically arises from technical fluctuations in the laser intensity incident on the photodetector, caused primarily by the relative intensity noise (RIN) of the laser source, as well as by other technical factors. Polarimetric detection techniques for magnetic resonance have proven effective in addressing the problem of suppressing such noise (e.g., see \cite{Budker2002,Li2008,Yang2024}). In the configuration considered, the suppression of intensity noise necessitates equalization of the resonance backgrounds (``pedestals'') depicted in Fig.~\ref{fig:4}(a), as the intensity noise is proportional to the magnitude of these pedestals. This is achieved through the use of a half-wave ($\lambda/2$) plate positioned before the WP (see Sec.~\ref{sec:31}). This plate serves to redistribute the optical power between the pump and probe beams such that the background contribution to the magnetic-resonance signal in the differential output of the balanced photodetector is nullified.

Consequently, the resonance line shape recorded in the differential channel of the BPD assumes a different form than that described by the analytical expressions (\ref{Winc})--(\ref{Woutp}). Omitting the routine derivations, the in-phase and quadrature signals, $I^{(in)}_{diff}$ and $I^{(out)}_{diff}$, as detected by the LIA from the differential channel of the BPD, can be expressed as follows:
 
%\begin{eqnarray}
%    &&I^{(in)}_{diff} = \alpha_0 \ell \xi I_0^2 \cos^2(\epsilon-\pi/4)\nonumber\\
%    &&\qquad\qquad\times\cos(4\eta_1) \sin^2(\theta)\frac{\Delta}{4\big(\delta^2+\Delta^2\big)}\,,\\[5pt]
%    &&I^{(out)}_{diff} = \alpha_0 \ell \xi I_0^2 \cos^2(\epsilon-\pi/4)\nonumber\\ 
%    &&\qquad\qquad\times\cos(4\eta_1) \sin^2(\theta)\frac{\delta}{4\big(\delta^2+\Delta^2\big)}\,.
%\end{eqnarray}

\begin{align}
    I^{(in)}_{diff} &= \alpha_0 \ell \xi I_0^2 \cos^2(\epsilon-\pi/4)\nonumber\\
    &\quad\times\cos(4\eta_1) \sin^2(\theta)\frac{\Delta}{4\big(\delta_m^2+\Delta^2\big)}\,,\label{eq:Iindiff}\\[5pt]
    I^{(out)}_{diff} &= -\alpha_0 \ell \xi I_0^2 \cos^2(\epsilon-\pi/4)\nonumber\\ 
    &\quad\times\cos(4\eta_1) \sin^2(\theta)\frac{\delta_m}{4\big(\delta_m^2+\Delta^2\big)}\,.\label{eq:Ioutdiff}
\end{align}

\noindent Here, $\eta_1$ is the rotation angle of the $\lambda/2$ plate placed in front of the Wollaston prism, at which the DC background levels of the magnetic resonances shown in Figs.~\ref{fig:4}(a) are equal, so that the background is completely canceled at the differential output of the photodetector. It can be shown that this angle is given by

\begin{equation}\label{AngleEta}
    \tan(4\eta_1) = \frac{1}{\cos(2\epsilon)}\Bigg[ \frac{M_z^{(0)}(\theta)/\mathcal{M}}{\sin(2\epsilon)} - 1\Bigg]\,,
\end{equation}

\noindent with

\begin{equation}\label{M0z}
    M_z^{(0)}(\theta) \approx \frac{\mathcal{R}_0\cos^2(\theta)}{\Delta}\,\mathcal{M}\,.
\end{equation}

The behavior of the signals (\ref{eq:Iindiff}) and (\ref{eq:Ioutdiff}) is shown in Fig.~\ref{fig:4}(c). For the modeling, we set $\eta_1$$\,=\,$$\pi/16$, which approximately corresponds to the value used in the experiments (Sec.~\ref{sec:3}). It can be readily shown that the relationship between the resonance height, linewidth, and slope given by Eq.~(\ref{eq:Slope}) also holds for the functions in Eqs.~(\ref{eq:Iindiff}) and (\ref{eq:Ioutdiff}). Therefore, Eq. (\ref{eq:Sensitivity}) can also be used to estimate the sensitivity of the sensor based on the parameters of the signal at the differential output of the photodetector.

In the model presented here, we do not take into account some effects that can lead to shifts of the MR. Some of these effects also depend on the relative orientation of the light wave vector and the measured magnetic-field vector, resulting in so-called heading errors of the sensor~\cite{Oelsner2019}. In addition to causing resonance shifts, such effects can substantially distort the line shapes of the resonances observed in the $X$ and $Y$ outputs of the LIA \cite{Zhang2025}. Therefore, for practical applications of the proposed scheme under conditions involving changes in the direction of the magnetic-field vector, it is preferable to use the signal from the $R$ output of the LIA, which is calculated from the $X$ and $Y$ signals as follows:

\begin{align}\label{eq:Rdiffchan}
R &= \sqrt{\big\{I^{(in)}_{diff}\big\}^2 + \big\{I^{(out)}_{diff}\big\}^2\,}\nonumber\\[5pt]
& = \frac{\alpha_0 \ell \xi I_0^2 \cos^2(\epsilon-\pi/4) \cos(4\eta_1) \sin^2(\theta)}{4\sqrt{\delta_m^2+\Delta^2}}\,.
\end{align}

\noindent As shown in~\cite{Zhang2025}, the MR in the $R$ output is not affected by the phase shift induced by the AC light shift and therefore remains symmetric. The MR in the $R$ output has the Lorentzian-like shape and the same height as the resonance in $I_{diff}^{(in)}$, while its linewidth is larger by a factor of $\sqrt{3}$. In Sec.~\ref{sec:3}, we investigate the behavior of the resonance specifically in the $R$ output.

Based on Eqs.~(\ref{AngleEta}) and (\ref{eq:Rdiffchan}), it can be shown that the optimal ellipticity is

\begin{equation}\label{eq:OptEllip}
    \epsilon_{opt} = \frac{1}{2}\arccos\bigg(\sqrt{\big(\sqrt{5}-1\big)/2}\bigg) \approx \pi/8\,.
\end{equation}

 \noindent At this ellipticity, the resonance height reaches its maximum. Given this ellipticity, the probe-wave intensity at the vapor cell entrance is approximately $15\%$ of the total intensity of the light beam. To derive Eq.~(\ref{eq:OptEllip}), we assumed that $\mathcal{R}_0$$\,\approx\,$$\Gamma$ and $\theta$$\,\approx\,$$\pi/2$, which is close to the optimal experimental conditions. 

The expression (\ref{eq:Rdiffchan}) also reflects an operating zone (OZ) of the sensor. This region is defined by the values of $\theta$ for which the resonance height decreases by a factor of two when $\theta$ is varied from its optimum value of $\pi/2$. As follows from (\ref{AngleEta}), the angle $\eta_1$ also depends on $\theta$. Therefore, to determine the sensor OZ, we can consider the function:

\begin{align}\label{eq:DeadZone}
&{\rm OZ}(\theta) = \sqrt{3}\,\sin^2(\theta)\cos\big[4\eta_1(\theta)\big]\nonumber\\[5pt]
& = \frac{\sqrt{3}\,\sin^2(\theta)\big|\sin(4\epsilon)\big|}{\sqrt{\sin^2(4\epsilon)+4\big[\frac{\mathcal{R}_0}{\Delta}\cos^2(\theta)-\sin(2\epsilon)\big]^2}}\nonumber\\[5pt]
&\approx \frac{\sqrt{3}\,\sin^2(\theta)}{\sqrt{1+4\big[\cos^2(\theta)-\sqrt{2}\big]^2}}\,.
\end{align}

\begin{figure*}[!t]
\centering
\includegraphics[width=1\linewidth]{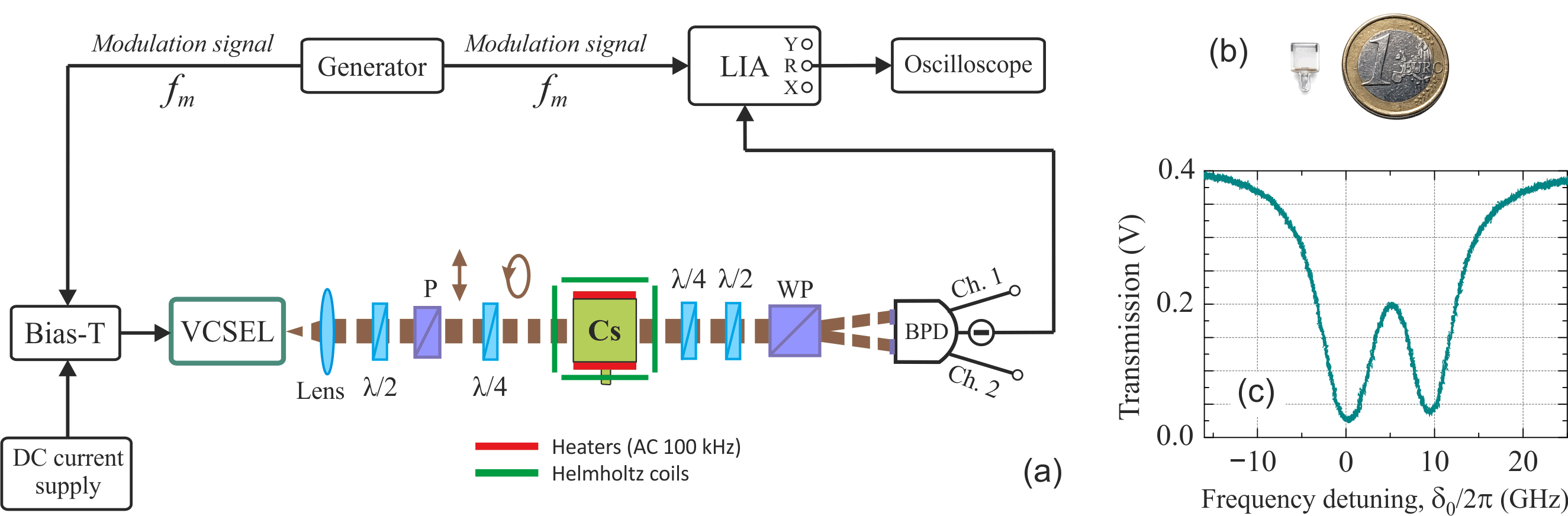}
\caption{\label{fig:5} (a) Experimental setup: VCSEL, vertical-cavity surface-emitting laser diode; P, polarizer; $\lambda$/2, $\lambda$/4, half-wave and quarter-wave plates, respectively; WP, Wollaston prism; BPD, balanced photodetector; LIA, lock-in amplifier. (b) Cs vapor cell next to a 1-euro coin. (c) The signal detected by the photodetector for a light beam transmitted through the cesium vapor cell at $T_{\rm cell}$$\,\approx\,$$85^\circ$C, $P_0$$\,\approx\,$$300$~$\mu$W, $\epsilon$$\,\approx\,$$45^\circ$, $B_0$$\,\approx\,$$50$~$\mu$T}
\end{figure*}

\begin{figure*}[!t]
\centering
\includegraphics[width=0.9\linewidth]{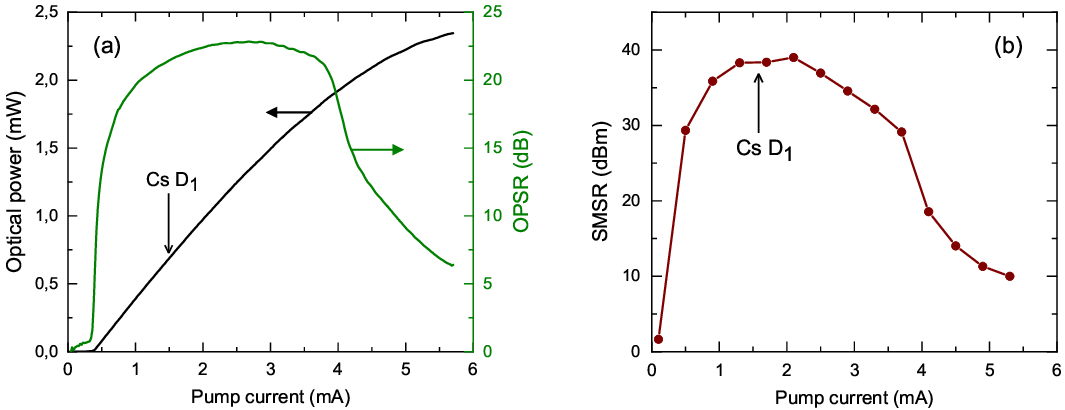}
\caption{\label{fig:6} VCSEL parameters as a function of the laser pump current: (a) Optical power and orthogonal polarization suppression ratio, and (b) Side-mode suppression ratio. The laser diode temperature is $\approx\,$$23^\circ$C}
\end{figure*}

\noindent Here, the factor $\sqrt{3}$ is introduced to normalize the function such that ${\rm OZ}(\pi/2)$$\,=\,$$1$. The approximate Eq.~(\ref{eq:DeadZone}) is derived under the assumptions: $\mathcal{R}_0$$\,\approx\,$$\Gamma$ and $\epsilon$$\,=\,$$\epsilon_{opt}$$\,\approx\,$$\pi/8$. Although Eq.~(\ref{eq:DeadZone}) exhibits a relatively complex dependence on $\theta$, in the most relevant angular range near $\pi/2$, where the sensor sensitivity remains acceptable, this function can be well approximated by the simple dependence ${\rm OZ}(\theta)$$\,\approx\,$$\sin^2(\theta)$.

It follows from Eq.~(\ref{eq:DeadZone}) that the operating zone of the sensor extends from $\theta$$\,\approx\,$$40^\circ$ to $\theta$$\,\approx\,$$140^\circ$ and $\theta$$\,\approx\,$$220^\circ$ to $\theta$$\,\approx\,$$320^\circ$. In other words, the sensor remains operational when the direction of the measured magnetic field deviates by up to $\pm$$50^\circ$ from the optimum orientation of $\theta$$\,=\,$$\pm$$\pi/2$. Note that the angular ranges outside these two intervals are conventionally referred to as the sensor dead zone. In particular, the MR vanishes entirely at $\theta$$\,=\,$$0,$$\,\pi$. This dead zone, corresponding to the configuration ${\bf B}$$||$${\bf k}$, is commonly referred to as a ``polar'' dead zone. It should be noted that in rf $M_x$ OPMs an additional dead zone is present, namely, there is an ``equatorial'' dead zone at $\theta$$\,=\,$$\pm$$\pi/2$ (e.g., see \cite{Schultze2012,Alexandrov2013}).

It should be noted that the analysis presented in this section is subject to certain simplifications. Specifically, expression (\ref{signal}) is valid only for the optically thin medium, where the absorption is assumed to be sufficiently small ($\alpha_0$$L_{\rm cell}$$\,\ll\,$$1$). Moreover, a more accurate expression should also account for the contribution of the quadrupole polarization moment in the atomic ground state~\cite{Castagna2011}. Nonetheless, for the purposes of a qualitative description of the physical mechanisms governing the formation of the magnetic resonance within the proposed experimental scheme, the present simplified treatment proves to be entirely adequate.

\section{Experiments}\label{sec:3}

\subsection{Experimental setup}\label{sec:31}

The experimental setup is shown in Fig.~\ref{fig:5}(a). The \mbox{VCSEL}, which was developed at the Ioffe Institute, is used in our experiments to excite MR in cesium vapor ($\lambda$$\,\approx\,$$894.6$~nm). The laser emission linewidth does not exceed $100$~MHz. Detailed information on the design features of the laser can be found in Ref.~\cite{Blokhin2019}. To ensure a large dynamic range of optical power and a relatively narrow emission linewidth (to suppress the linewidth broadening effect due to laser self-heating), the VCSEL with an operating temperature of $20$$-$$30^\circ$C is selected. The laser emission wavelength can be continuously varied by means of either the laser diode temperature or drive current, which are controlled using TED200C and LDC201CU controllers (Thorlabs), respectively. The laser is mounted using an LDM90 adapter (Thorlabs). As seen in Fig.~\ref{fig:6}(a), left $y$-axis, at a pump current of $\approx\,$$1.5$~mA and an operating temperature of $\approx\,$$23^\circ$C, the optical power reaches $700$~$\mu$W. Under these conditions, the VCSEL emission wavelength corresponds to the D$_1$ line of $^{133}$Cs. The laser emission is linearly polarized with an orthogonal mode suppression ratio (OPSR) of over $20$~dB (see Fig.~\ref{fig:6}, right $y$-axis). The laser exhibits single-mode operation with a side-mode suppression ratio (SMSR) of over $35$~dB [Fig.~\ref{fig:6}(b)].

%The low-frequency spectrum of the RIN has the appearance of 1/f n-noise with a smooth transition to white noise at frequencies above 50 kHz. As expected, the noise level decreases with increasing output optical power, but significantly exceeds the photodetector's shot noise (see Fig. 2). At a pump current of 1.5 mA and an operating temperature of 23 °C, the VCSEL emission wavelength corresponds to the D1 line of 133Cs, the optical power reaches 700 μW, the emission linewidth does not exceed 100 MHz, noise at a frequency of more than 100 kHz does not exceed -140 dB/Hz.

\begin{figure*}[!t]
\centering
\includegraphics[width=0.9\linewidth]{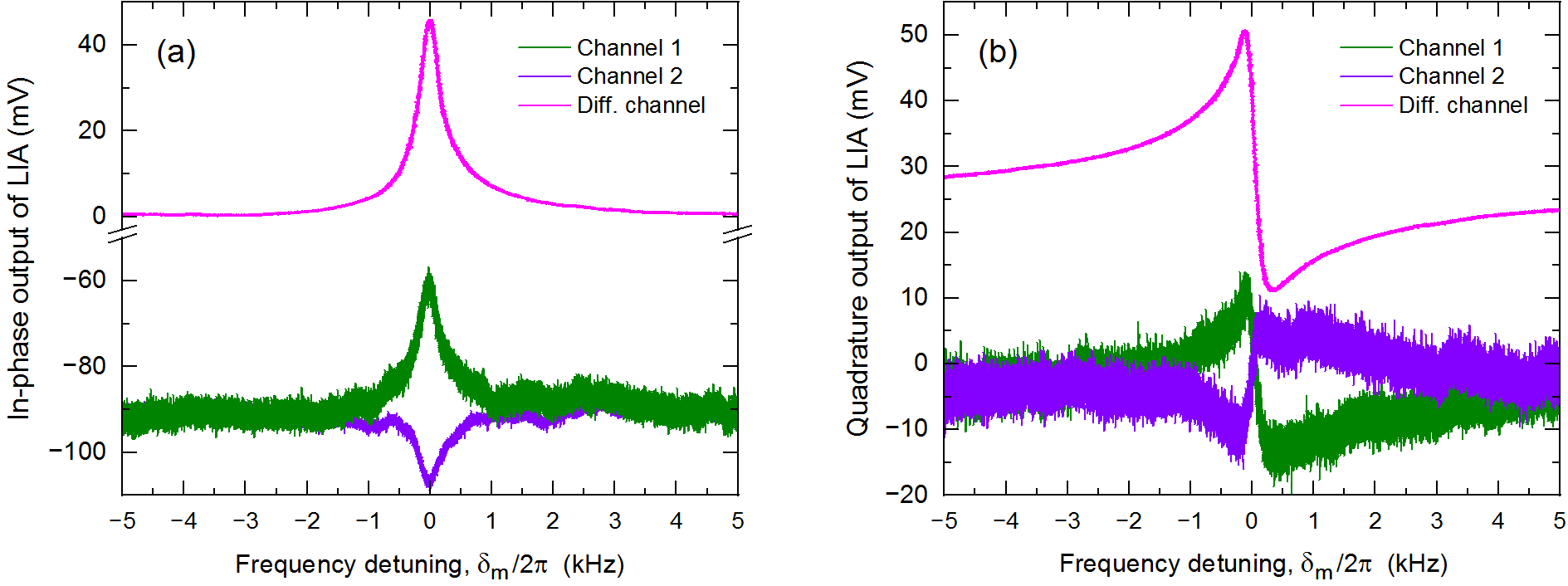}
\caption{\label{fig:7} (a) Magnetic resonance curves obtained at the in-phase output of the LIA, when its input is switched between channel~1 (green curve), channel~2 (violet), and the differential channel (pink) of the BPD. (b) The same resonance obtained at the quadrature output of the LIA. The experimental conditions: $T_{\rm cell}$$\,\approx\,$$85^\circ$C, $P_0$$\,\approx\,$$350$~$\mu$W, $\epsilon$$\,\approx\,$$25^\circ$, LIA time constant is $300$~$\mu$s}
\end{figure*}

In our experiments, the laser emission wavelength is controlled by a WS7 wavelength meter (Angstrom Co. Ltd.). It is tuned close to the center of the absorption line corresponding to dipole transitions $F_g$$\,=\,$$4$$\,\to\,$$F_e$$\,=\,$$3,$$\,4$ [left dip in Fig.~\ref{fig:5}(c)]. Here, $F_g$ and $F_e$ are the total angular momenta in the atomic ground and excited states, respectively. Due to the high pressure of the buffer gas in the cell ($\approx\,$$200$~Torr), the neighboring optical transitions $F_g$$\,=\,$$4$$\,\to\,$$F_e$$\,=\,$$3$ and $F_g$$\,=\,$$4$$\,\to\,$$F_e$$\,=\,$$4$ are not spectrally resolved.

After passing through the lens, the laser beam has a $1/$$e^2$ diameter of $\approx\,$$1.7$~mm. The radiation then passes through a half-wave plate and a polarizer, which serve to adjust the optical power of the beam. The optical power at the cell entrance is controlled by a PM100A power meter (Thorlabs). The laser beam then passes through a quarter-wave plate, acquiring elliptical polarization. The ellipticity parameter, $\epsilon$, coincides with the rotation angle of the plate. In the vapor cell, the different absorption coefficients experienced by the circularly polarized probe and pump components of the light beam, a phenomenon known as circular dichroism, result in a change in the ellipticity of the light relative to its initial value. To measure this change, the beam is directed to a polarimeter consisting of a $\lambda/4$ plate, a $\lambda/2$ plate, a Wollaston prism (WP), and a balanced photodetector (PDB210A, Thorlabs). After passing through the $\lambda/4$ plate, the circularly polarized pump and probe waves are converted into mutually orthogonal linear polarizations. These waves can therefore be separated by the WP, which directs the pump wave to one input of the balanced photodetector and the probe wave to the other. The $\lambda/2$ plate is used to redistribute the optical power between the pump and probe components of the beam, as discussed below.

A $5$$\times$$5$$\times$$5$~mm$^3$ cubic borosilicate-glass cell filled with Cs atoms is used in the experiments [Fig.~\ref{fig:5}(b)]. The cell additionally contains a buffer gas mixture composed of Ar and N$_2$ in a ratio of approximately $80$$:$$120$. The cell is heated by thin resistive elements mounted on its side and front faces. Heating is realized with AC current at $100$~kHz, which has no visible effect on the resonance linewidth and height. The sensor performance is characterized inside a three-layer magnetic shield providing a residual magnetic field at the cell of no more than $50$~nT. Two pairs of Helmholtz coils are employed to produce a magnetic field within the Earth's range, oriented in the $xz$-plane (see Fig.~\ref{fig:1}). A DC voltage is applied to the coils using a DG4162 function waveform generator (Rigol).

%The bandwidth is limited by the photodetector and amounts to $\approx\,$$1$~MHz
%, $f_m$ sweeping rate is $10$~Hz

%To mitigate magnetic noise and evaluate the measurement sensitivity, the vapor cell and the Helmholtz coils are housed within a three-layer magnetic shield.

The laser injection current is modulated at a frequency $f_m$ close to the Larmor frequency $\Omega_L$$\,=\,$$2\pi$$\times$$160$~kHz using a function waveform generator DG2102 (Rigol) and a bias tee, which is integrated into the LDM90 adapter. This Larmor frequency corresponds to a magnetic field of $B_0$$\,\approx\,$$46$~$\mu$T. For MR observation, $f_m$ is slowly swept at a rate of $\approx\,$$10$~Hz. For all experiments other than operating-zone measurements, the magnetic field is oriented perpendicular to the light wave vector, i.e., along the $x$-axis. Sweeping $f_m$ near $\Omega_L$ leads to a sharp change in the ellipticity parameter of the light beam at the exit of the cell due to MR. This change is observed using a polarimeter and an SR865A lock-in amplifier (Stanford Research Systems), with its output signal monitored on a DHO4204 oscilloscope (Rigol). The noise voltage spectral density is measured using an SR1 spectrum analyzer (Stanford Research Systems).

%The same oscilloscope is used to analyze the resonance voltage noise via the fast Fourier transform (FFT) option.

%The same adapter is also used to inject, from a DG2102 function generator (Rigol), a sinusoidal modulation signal at frequency $f_m$ into the laser drive current, thus producing frequency modulation of the optical carrier. The identical signal is simultaneously delivered to the LIA (SR865A, Stanford Research Systems) as the reference input.

\subsection{Measurements}\label{sec:32}

\begin{figure*}[!t]
\centering
\includegraphics[width=1\linewidth]{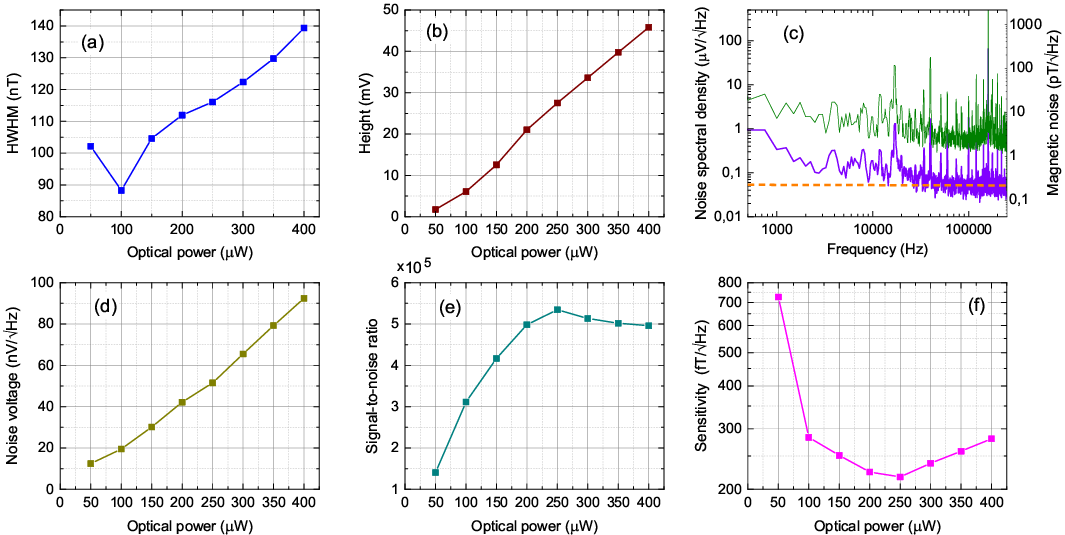}
\caption{\label{fig:8} (a), (b), (d)-(f) Magnetic-resonance parameters as a function of the optical power incident on the vapor cell: half width at half maximum (a); height (b); noise voltage spectral density at $f_m$$\,=\,$$160$~kHz for the proposed polarimetric scheme (d); SNR in a 1-Hz bandwidth (e); and sensor sensitivity (f). (c) Comparison of the noise voltage spectral density obtained using the conventional single-beam Bell-Bloom scheme with circularly polarized light (green) and the proposed polarimetric scheme with elliptically polarized light (violet) at $P_0$$\,\approx\,$$250$~$\mu$W. Measurement parameters: $\epsilon$$\,\approx\,$$25^\circ$, $T_{\rm cell}$$\,\approx\,$$85^\circ$C, and LIA time constant of $100$~$\mu$s. Solid lines in panels a, b, d, e, and f are drawn as a guide to the eye}
\end{figure*}

Fig.~\ref{fig:7} illustrates the magnetic resonances obtained from channel~1, channel~2, and differential channel of the BPD. In accordance with theoretical predictions (Sec.~\ref{sec:23}), the resonances in channels 1 and 2 exhibit opposite signs. Even without a direct measurement of the noise voltage, it becomes evident that the resonance in the differential channel is considerably less noisy than that in each of the individual channels. As seen from Fig.~\ref{fig:7}(a), at the output of each of channels 1 and 2, the magnetic resonances exhibit a background signal of $\approx\,$$-90$~mV. This means that the amplitude of the modulated response of the medium in each channel is $\approx\,$$90$~mV, while the peak-to-peak modulation amplitude is $\approx\,$$180$~mV. Since, in the balanced-channel configuration, the optical power is distributed approximately equally between the two channels by means of the $\lambda/2$ plate [see Fig.~\ref{fig:5}(a)], the peak-to-peak modulation amplitude of the light beam before the Wollaston prism is therefore approximately $180\times2$$\,=\,$$360$~mV. This value is consistent with the magnitude of the left absorption dip shown in Fig.~\ref{fig:5}(c).

To determine the optimal optical power for exciting magnetic resonance, a series of measurements of the resonance parameters (linewidth, height, and signal-to-noise ratio) is performed, and the magnetic-field measurement sensitivity is estimated based on the results. In particular, Fig.~\ref{fig:8}(a) reflects the behavior of the resonance linewidth. At low optical powers, the resonance exhibits a slight narrowing with increasing power, whereas at higher power levels, conventional linear power broadening is observed. Such narrowing can be ascribed to the light-induced narrowing effect \cite{Savukov2005}, which is known to manifest in various magneto-optical experiments \cite{Brazhnikov2026,Bevington2022,Chen2024}, as long as radiation trapping in the atomic vapor remains negligible \cite{Han2017}. In contrast to the resonance width, its height exhibits almost linear increase with optical power [Fig.~\ref{fig:8}(b)]. Only in the regime of low optical powers, $P_0$$\,\lesssim\,$$200$~$\mu$W, does the dependence exhibit a nonlinear behavior, as is evident from the analytical expression (\ref{eq:Rdiffchan}).

%In the classical Bell--Bloom AM and FM schemes, the magnetic resonance is detected via the intensity of the light beam transmitted through the vapor cell \cite{Bell1961,Martinez2010}. In such schemes, laser intensity noise typically constitutes the most significant contribution to the overall resonance noise (a comparative analysis of various noise types in atomic magnetometry is presented in Ref.~\cite{Krzyzewski2019}). Therefore, polarimetric methods for magnetic resonance detection are widely employed in atomic magnetometry, as they enable a substantial improvement in the signal-to-noise ratio and, consequently, in the measurement sensitivity.

Fig.~\ref{fig:8}(c) demonstrates the effect of intensity noise suppression. The upper trace (green) shows the noise voltage spectral density, $\mathcal{N}$, on the MR slope, i.e., it is recorded at $\delta_m$$\,\approx\,$${\rm HWHM}$ to account for frequency-to-amplitude noise conversion. This measurement is carried out in the classic Bell--Bloom scheme with a circularly polarized beam ($\epsilon$$\,\approx\,$$\pi/4$), using a single photodetector channel. Similar measurement is performed using the balanced output of the photodetector (bottom trace), whose signal is proportional to a change in the light polarization ellipticity rather than to the total optical intensity. In the figure, a sharp spike at $f$$\,\approx\,$$160$~kHz corresponds to the light-field modulation signal. In the figure, the magnetic noise spectral density shown on the right $Y$-axis is obtained from the magnetic resonance slope. A comparison of the green and violet curves in Fig.~\ref{fig:8}(c) reveals that the polarimetric MR detection suppresses the noise by more than an order of magnitude. It should be noted that the detection scheme used enables reduction of not only the intensity noise originating from the laser source, but also its polarization noise, because the latter being converted into intensity noise by a polarizer [see Fig.~\ref{fig:5}(a)].

To characterize the dependence of $\mathcal{N}$ measured at the balanced output of the photodetector on the optical power, the noise voltage is evaluated at a frequency near the $160$~kHz Larmor frequency. In Fig.~\ref{fig:8}(c), this value is approximately $50$~nV/$\surd$Hz, as schematically indicated by the dashed orange line. As shown in Fig.~\ref{fig:8}(d), $\mathcal{N}$ increases gradually with light intensity, following a linear-like dependence. Such a dependence implies that the observed noise is still not of a shot-noise nature. Several mechanisms may be responsible for this linear scaling. The first concerns the imperfect common-mode rejection between the two photodetector channels, which results in incomplete subtraction of the laser intensity noise. This imperfection can be caused, in particular, by imprecise orientation of the half-wave plate in front of the Wollaston prism. The second mechanism involves polarization noise of the light beam, which may stem from a variety of effects, ranging from fundamental atomic spin fluctuations to vibrations and parasitic interference within the optical components of the setup (for a discussion, see, e.g., Refs.~\cite{Shah2009,Novikova2015,Tian2025}). These polarization fluctuations are anticorrelated in the two detection channels of the BPD, so that they are not canceled at the balanced output. Many of these noise contributions, however, could be further suppressed by building a more compact and rigid sensor head. Of particular interest in this regard is the application of miniaturized integrated polarimeters, akin to the one used in Ref.~\cite{Cho2026}.

%In our setup, this misalignment could be reduced by employing another half-wave plate mount of higher precision, equipped with a micrometer.

As follows from Fig.~\ref{fig:8}(e), the $1$-Hz-bandwidth SNR attains quite high values, reaching $\approx\,$$5$$\times$$10^5$. Using this parameter, the sensitivity of the sensor, $\delta B$, can be estimated from Eq.~(\ref{eq:Sensitivity}). The dependence of $\delta B$ on optical power is shown in Fig.~\ref{fig:8}(f). The highest sensitivity, i.e. the lowest $\delta B$, is achieved at $P_0$$\,\approx\,$$250$~$\mu$W, and amounts to $\approx\,$$220$~fT/$\surd$Hz. This optical power, of course, is optimal under the given conditions.

\begin{figure*}[!t]
\centering
\includegraphics[width=0.9\linewidth]{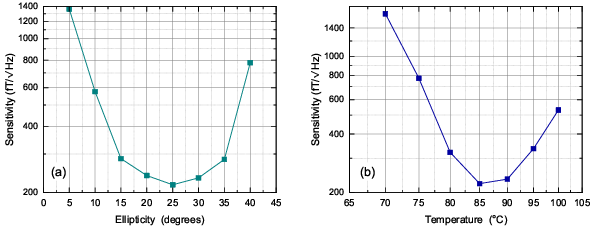}
\caption{\label{fig:9} Dependence of the sensor sensitivity on (a) the ellipticity of the light polarization and (b) the temperature of the atomic vapor in the cell. $P_0$$\,=\,$$250$~$\mu$W for both plots. $T_{\rm cell}$$\,=\,$$85^\circ$C in (a) and $\epsilon$$\,=\,$$25^\circ$ in (b). Solid lines serve as a guide to the eye}
\end{figure*}

In addition to the measured sensitivity under the current experimental conditions, the ultimate sensitivity of a quantum sensor is commonly characterized by assuming that a certain quantum limit is reached \cite{Budker2007,Fabricant2023}. Two types of quantum noise are commonly considered when estimating the ultimate sensitivity of a quantum sensor: spin-projection noise (SPN) and photon shot noise (PSN). A more detailed discussion of this issue can be found in a recent work~\cite{Lei2025}. In particular, SPN imposes the following limit:

\begin{equation}
    \delta B_{\rm SPN} \approx \frac{1}{\gamma\sqrt{N_{\rm at}\tau}}\,,
\end{equation}

\noindent where $N_{\rm at}$ is the number of cesium atoms within the active sensing volume, $\tau$$\,=\,$$1$$/$$2\pi\Gamma$ is the spin coherence time. Based on formula from Ref.~\cite{SteckCs}, under the conditions used in our experiments, the resulting estimate of the sensitivity in the SPN limit is $\delta$$B_{\rm SPN}$$\,\approx\,$$2.5$~fT/$\surd$Hz.

To estimate the PSN-limited sensitivity, we first calculate the SNR in a 1-Hz bandwidth, which is given by

\begin{equation}\label{eq:SNRPSN}
    {\rm SNR}_{\rm PSN} \approx \frac{I_{\rm ph}}{\sqrt{2eI_{\rm ph}}} = \sqrt{\frac{\varkappa N_{\rm ph}}{2}}\,,
\end{equation}

\noindent where $I_{\rm ph}$$\,=\,$$\varkappa$$e$$N_{\rm ph}$ is the photocurrent, with $\varkappa$ denoting the quantum efficiency of the photodiode ($\approx\,$$0.8$), $e$ the elementary charge, and $N_{\rm ph}$ the photon flux, i.e., the number of detected photons per $1$~s, which is given by $N_{\rm ph}$$\,=\,$$P_0$$\lambda$$/$$2\pi\hbar$$c$. Note that Eq.~(\ref{eq:SNRPSN}) assumes Poissonian statistics of the detected photons. Other states of the light field, such as squeezed states, lead to different expressions; however, they are not considered here. Based on the background voltage values in channels 1 and 2 shown in Fig.~\ref{fig:7}(a), we estimate the average total optical power incident on the two photodiodes of the photodetector to be approximately $25$~$\mu$W. This corresponds to a photon flux of $\approx\,$$10^{14}$~s$^{-1}$, and we immediately get ${\rm SNR}_{\rm PSN}$$\,\approx\,$$6.7$$\times$$10^6$. The sensitivity can be estimated from Eq.~(\ref{eq:Sensitivity}), so that we obtain $\delta$$B_{\rm PSN}$$\,\approx\,$$17$~fT/$\surd$Hz.

\begin{figure}[!b]
\centering
\includegraphics[width=1\linewidth]{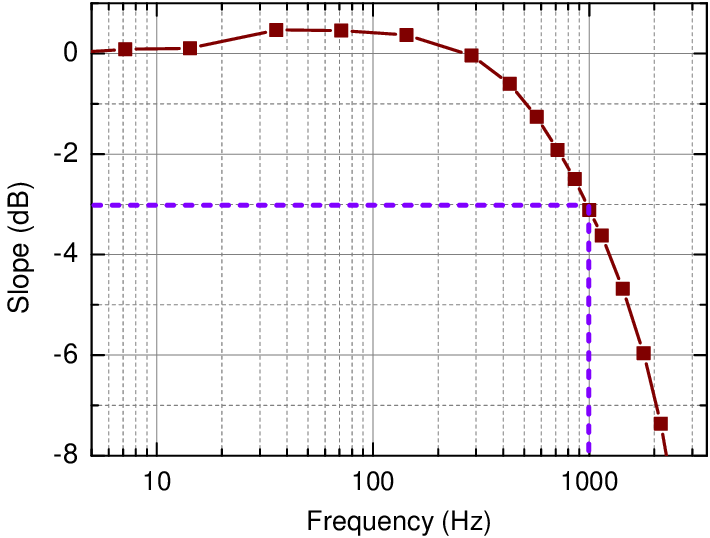}
\caption{\label{fig:10} Magnetic-resonance slope in the quadrature channel of the LIA as a function of the frequency-sweep rate. Parameters: $P_0$$\,\approx\,$$320$~$\mu$W, $T_{\rm cell}$$\,\approx\,$$85^\circ$C, $\epsilon$$\,\approx\,$$25^\circ$, LIA time constant is 100~$\mu$s}
\end{figure}

Thus, we conclude that the fundamental factor primarily limiting the sensitivity in our scheme is the PSN limit. In general, the estimates presented above indicate that, after eliminating technical noise sources, the proposed approach could achieve a sensitivity comparable to that of state-of-the-art compact SERF OPMs used for noninvasive medical diagnostics, which typically exhibit noise floors of approximately $10$$-$$40$~fT/$\surd$Hz \cite{QuSpin,Gutteling2023}.

In addition to optical power, another key parameter in the investigated scheme is the ellipticity of the light polarization at the vapor cell entrance. Figure~\ref{fig:9}(a) presents the dependence of the sensor sensitivity on the ellipticity, from which it follows that the optimal ellipticity is approximately $25^\circ$. This value is in adequate quantitative agreement with that predicted by the simplified theoretical model developed in Sec.~\ref{sec:23}.
 
Figure~\ref{fig:9}(b) shows that the optimal vapor temperature is approximately $85^\circ$C. At this temperature, the light transmission at the center of the spectral line of the $F_g$$\,=\,$$4$$\,\to\,$$F_e$$\,=\,$$3,$$\,4$ transitions is $\approx\,$$8\%$ [see Fig.~\ref{fig:5}(c)]. The corresponding optical depth (OD) is $\approx\,$$2.5$. We note that the theoretical model (Sec.~\ref{sec:2}) has been developed for an optically thin medium, ${\rm OD}$$\,\ll\,$$1$, which allowed us to obtain an analytical solution of the Bloch equation. Therefore, quantitative deviations between the theoretical and experimental results may be partly attributed to the violation of this condition in the experiments.

The bandwidth of a magnetic sensor characterizes the maximum rate of magnetic-field variation that can be resolved while maintaining at least half of the optimum sensitivity. The most straightforward way to measure this characteristic is to use the magnetic-resonance signal at the $Y$ output of the LIA, where the resonance appears as an error-like signal. The bandwidth of the sensor can then be determined from the twofold ($-3$~dB) reduction in the slope of this signal as the FM sweep rate ($f_s$) is increased. Thus, when measuring the sensor bandwidth, the light field modulation frequency is varied according to the following law:

\begin{equation}
f_m(t) = f_{m0} + \Delta_{m}\cos(f_st)\,.    
\end{equation}

\noindent Here, $f_{m0}$$\,=\,$$160$~kHz is the Larmor frequency, $\Delta_{m}$$\,=\,$$100$~Hz is the frequency deviation (corresponding to approximately one-tenth of the resonance linewidth), while $f_s$ ranges from $5$~Hz to several kilohertz. Figure~\ref{fig:10} shows the resonance slope (normalized to its value at $f_s$$\,=\,$$5$~Hz and expressed in dB) as a function of $f_s$. As seen from the figure, the resonance slope decreases by $3$~dB at $f_s$$\,\approx\,$$1$~kHz. We note that this value is in qualitative agreement with the linewidth of the magnetic resonance itself, which is $\approx\,$$0.9$~kHz at the given optical power.

\begin{figure}[!t]
\centering
\includegraphics[width=1\linewidth]{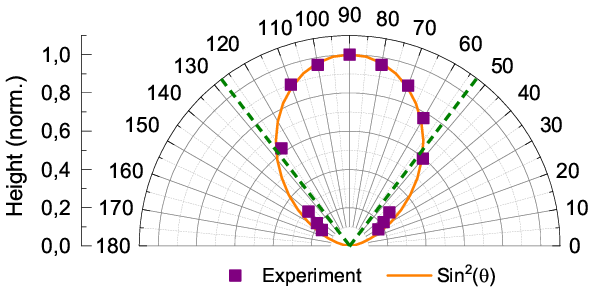}
\caption{\label{fig:11} Measurement of the sensor dead zone. The green lines indicate the values of the angle $\theta$ at which the resonance height decreases by a factor of two relative to that in the optimal configuration, corresponding to $\theta$$\,=\,$$90^\circ$. Measurement parameters: $P_0$$\,\approx\,$$300$~$\mu$W, $T_{\rm cell}$$\,\approx\,$$85^\circ$C, $\epsilon$$\,\approx\,$$25^\circ$, LIA time constant is 100~$\mu$s}
\end{figure}

Finally, we have measured the range of angles $\theta$ over which the sensor sensitivity decreases by no more than a factor of two. As the sensor is tilted away from its optimum orientation ($\theta$$\,=\,$$90^\circ$), the degradation in sensitivity is primarily caused by a reduction in the magnetic-resonance height. As seen in Fig.~\ref{fig:11}, the sensor operating range extends to approximately $\pm\,$$40^\circ$ from the optimum orientation. This range is somewhat narrower than that predicted by the theoretical model presented in Sec.~\ref{sec:23}. Nevertheless, as shown in the figure, for small and moderate deviations from $\theta$$\,=\,$$90^\circ$, the operating range is well described by the simple $\sin^2(\theta)$ dependence, in qualitative agreement with Eq.~(\ref{eq:DeadZone}).

\section{Conclusions}\label{sec:4}

We have proposed and investigated an optically pumped atomic magnetometer scheme based on a single laser beam with elliptical polarization. The magnetic resonance in our scheme arises from circular dichroism and is detected through a change in the ellipticity parameter. As in other polarimetric techniques, the proposed detection approach substantially suppresses laser-intensity noise and enables a high magnetic-field sensitivity of approximately $220$~fT/$\surd$Hz under the current experimental conditions. The sensor exhibits only a single dead zone, while maintaining a bandwidth on the order of $1$~kHz. The relatively simple sensor configuration and the demonstrated performance make the proposed approach attractive for a broad range of geophysical applications, including geological surveying and UAV-based archaeological exploration.

The demonstrated sensitivity is not yet limited by fundamental quantum noise, such as photon shot noise. With further suppression of technical noise sources, a sensitivity close to $20$~fT/$\surd$Hz appears feasible. For example, cyclic heating of the vapor cell~\cite{Savukov2016,Wang2024} could reduce the electromagnetic noise generated by the heating elements, while an integrated polarimeter design~\cite{Sebbag2021,Cho2026} could substantially suppress the effects of polarization noise. To date, our sensor has been tested with insufficient shielding from external magnetic fields, which could be further improved in future implementations. 

Moreover, in addition to reducing technical noise, the sensitivity could be further improved by optimizing the magnetic-resonance parameters. For example, we expect that the use of square-wave frequency modulation \cite{Martinez2010,Gerginov2017} could further increase the resonance height and, consequently, improve the sensitivity of our scheme. Rather than applying the frequency modulation at the slope of the absorption line, as in our experiments, the modulation can be applied across the entire absorption profile, with the signal detected at the second harmonic of the modulation frequency \cite{Martinez2010}. This approach could further improve both the signal-to-noise ratio and the sensitivity. In the latter case, however, it is preferable to use a cell with a higher buffer-gas pressure so that the absorption profile forms a single broadened line without resolved ground-state hyperfine structure in the atom, in contrast to the conditions used in our experiments.

Further improvement in sensitivity could broaden the range of applications of the proposed atomic magnetometry scheme, particularly in noninvasive medical diagnostics, including magnetocardiography (MCG) and magnetoencephalography (MEG). At the same time, in contrast to SERF OPMs, which are currently being actively developed for biomedical applications, the proposed sensor is not restricted to operation in the zero-field environments. In particular, a gradiometric configuration could enable its operation in the Earth's magnetic field -- an important capability for the development of magnetic sensors for biomedical applications~\cite{Brisinda2023,Petrenko2026}. One more important feature of the approach considered here is that the cell temperature is substantially lower than that in alkali-metal-based SERF OMPs. This may both extend the sensor lifetime and mitigate requirements to thermal isolation of the sensor housing from the heated vapor cell, which is crucial for medical applications. However, a more comprehensive investigation is required to assess the feasibility and performance of this approach under realistic experimental conditions.

%\vspace{20mm}

%About miniaturization: PPOP scheme or two-VCSEL scheme, microfabricated polarimeter (Yang: Single-beam pump-probe NMOR...).
%About improving sensitivity by meander and simmetric modulation of the absorption line.

\begin{acknowledgments}
The work was supported by the Russian Science Foundation (Grant No. 26-12-00132). The development of the VCSEL was supported by the Ministry of Science and Higher Education of the Russian Federation within the framework of the state assignment of the Ioffe Physical-Technical Institute (Project Code: FFUG-2025-0006).
\end{acknowledgments}

%\begin{widetext}

%\begin{equation}
%\frac{dI_p}{dz} = -\frac{3\gamma n_a \lambda^2}{8\pi\gamma_{eg}}
%\Biggl(1-\frac{2\chi_p\Gamma\Delta}{\Delta^2+\delta_R^2}\Bigl[1+\frac{\chi_c}{\chi_p}\cos{\psi} - \frac{\chi_c\delta_R}{\chi_p\Delta}\sin{\psi}\Bigr]\Biggr)\times I_p\,.
%\end{equation}

%\end{widetext}

%\begin{thebibliography}{89}

%\bibitem{Alzetta1976}
%A.~Alzetta, A.~Gozzini, L.~Moi, and G.~Orriols, An experimental method for the observation of r.f. transitions and laser beat resonances in oriented Na vapour, Nuovo Cimento B \textbf{36}, 5 (1976).

%\end{thebibliography}

%\bibliographystyle{unsrtnat}

\bibliography{references}

\end{document}